# Revising the Wilks Scoring System for pro RAW Powerlifting

22.3.2019

Anna Khudayarov, MD

Emergency Medicine, University of Turku and Turku University Hospital

**Abstract**

Purpose: In powerlifting the total result is highly dependent on the athlete's bodyweight. Powerlifting is divided to equipped and RAW types. Pro RAW powerlifting competitions use the Wilks scoring system to compare and rank powerlifting results across bodyweights, to choose the winners. The Wilks formulas are $5^{th}$ order polynomials fitted to equipped powerlifting results from the years 1987 – 1994. The Wilks formulas were meant to be updated every 2 to 5 years, but have never been revised. This study aims to update the Wilks scoring system for use in pro RAW powerlifting.

Methods: The 10 highest RAW powerlifting totals per weight class for bodyweights 60 – 175 kg for men, and 44 kg and up for women, were collected from the openpowerlifting.org database. Polynomials were fitted to these data separately for men and women, using polynomial regression. The scores for men and women were calculated using these fits, and the results are assessed.

Results: $4^{th}$ order polynomials were chosen for best describing the dependence of powerlifting total result on bodyweight. The men's scores were normalized to 500 points, and women's to 455 points, on those fitted curves, to bring the top scores to the same level across weight classes of each sex. These scores appeared similar across the different weight classes for both men and women. This even scoring across body weights is here considered the single most important characteristic of a scoring system for pro competitions.

**Key words:** Wilks score, revised prediction curve, polynomial regression, powerlifting

**Background**

Powerlifting is a sport combining squat, bench press, and deadlift. A powerlifting competition result is the total of these 3 lifting forms. Powerlifting is divided into equipped (using supporting suits) and RAW powerlifting (no suits). RAW powerlifting can be done with or without knee wraps (here called simply RAW powerlifting, with both forms included). In contrast to weightlifting, which is an Olympic sport, there is a clear difference between amateur powerlifting and pro powerlifting that has invitational money prize competitions. The US Open pro RAW powerlifting competition with over 120 000 USD prize money can be considered the gold standard in pro powerlifting.

The powerlifting total is highly dependent on an athlete's bodyweight. Various scoring methods have been developed to compare athletes across different bodyweights. Such scoring aims to be fair and unbiased, by giving every weight class an equal possibility to win a competition. There are independently adopted coefficients to correct for age, for teen and master age classes, which can be used together with all the different scoring systems. The scores are generally "calibrated" by fitting a function of bodyweight to a sampling of results, and are used by dividing the total in kilos by this function of bodyweight, like a correction factor. Most often used function types in adjusting strength by bodyweight are polynomial, power function, and loglinear (1, 2, 3). Methods taking into account the 3 different events that contribute to the powerlifting total have also been suggested (2, 5). Currently the method adopted by the amateur powerlifting federation IPF uses a loglinear function (3, 4). The current system of the IPF was introduced to use on January 1$^{st}$, 2019 (6). The prediction curve was fitted to an amateur result sampling of almost 30 000 results (3), and thus works well in amateur competitions, but it fails to serve fairly in pro level competition results. Therefore, the pro RAW powerlifting competitions use the Wilks scoring system.

However, the Wilks formula was made by fitting a 5$^{th}$ order polynomial to IPF's equipped powerlifting results from 1987 to 1994 (unpublished data), so it has had an opportunity to age. The sample size has been clearly smaller than used for the new IPF score, and thus worked well in pro level competitions. The Wilks scoring system was shown to be unbiased with IPF's World Championship results for years 1996 and 1997 (7).

The Wilks formula was meant to be updated every 2 to 5 years, but has, in fact, never been updated. The Wilks scoring system is nowadays used in pro RAW powerlifting competitions, which is far from the amateur equipped powerlifting for which the system was created. The Wilks formulas favour heavy weight men and middle weight women. Among the World's current 10 top ranked athletes, according to Wilks scores, there are no men below 100 kg weight class; for women 6 out of the 10 are middleweight (7). There is a clear need to update this system.

This study aimed to revise or update the Wilks formulas for both men and women, by using the All-Time top ranked RAW powerlifting results from the openpowerlifting.org database. The formulas are intended for use in pro competitions, and thus only pro level results will be used to fitting the prediction curves. Our aim is to produce a score that would distribute the top 10 ranked in the World evenly across all weight classes.

**Methods**

The 10 all-time top ranked RAW powerlifting results per every traditional weight class (-44 kg, -48 kg, -52 kg, -56 kg, -60 kg, -67,5 kg, -75 kg, -82,5 kg, -90 kg, -100 kg, -120 kg, -140 kg, and +140 kg for men; and -44 kg, -48 kg, -52 kg, -56 kg, -60 kg, -67,5 kg, -75 kg, -82,5 kg, -90 kg, and + 90 kg for women) were collected from the openpowerlifting.org on February 27$^{th}$, 2019. The 10 per weight class were chosen to produce a "high achiever" curve, as opposed to using all the powerlifting results.

The two missing body weights, both from the kg to kg Big Dogs pro competition, were replaced with the latest previous competition weights of the same athletes, checked from their athlete files in openpowerlifting.org.

The men under 60 kg and women under 44 kg classes were omitted from the analysis (aside from a couple of the very highest results in these weight classes, to help extrapolate the prediction curve), because of the methodological problem that polynomial correction overestimates the comparatively large variance of scores in the very lightweight classes (i.e., the data are by their nature heteroscedastic). Notably, there has been no competitor men under 60 kg or lighter, or women under 44 kg, in the US Open pro competitions, which can be considered the gold standard among pro RAW powerlifting competitions. Also, men over 175 kg were omitted from the analysis, because their results were clearly below those of the lighter heavyweight athletes. Including them would have produced a curve with scores that decrease with weight, while in principle the weight limit of a class serves as an upper limit to bodyweight, which necessarily gives a steadily increasing scoring curve. The biased scores are shown using men's data and 4$^{th}$ order polynomial prediction curve in the Appendix (figures 7A and 7B).

**Data analysis**

The current Wilks prediction curve was plotted along with the RAW powerlifting results by body weight, and the residuals were calculated to evaluate the goodness of fit and whether revising the fit would be necessary.

$2^{nd}$, $3^{rd}$, $4^{th}$ and $5^{th}$ order polynomials were fit separately to men's and women's RAW powerlifting results, and compared for best depicting the distribution of the results. $4^{th}$ order polynomials were chosen for further testing.

The men's scores were normalized to 500, and women's to 455 points, to bring the top scores to the same level across weight classes of each sex. Diagnostic plots of the scores were produced to check quality of the fits. A horizontal trend line of the scores was considered fair, or unbiased by bodyweight. The top 10 scores should fall evenly to the weight classes.

The polynomial regression analyses were done using the freely downloadable Jamovi software on a Windows PC. The plots were created in Microsoft Excel.

**Results**

The Wilks prediction curves did not follow the physiological distribution in the RAW powerlifting data (figures 1A-D). The Wilks scoring system produced clearly biased distributions for both men and women (figures 2A and 2B).

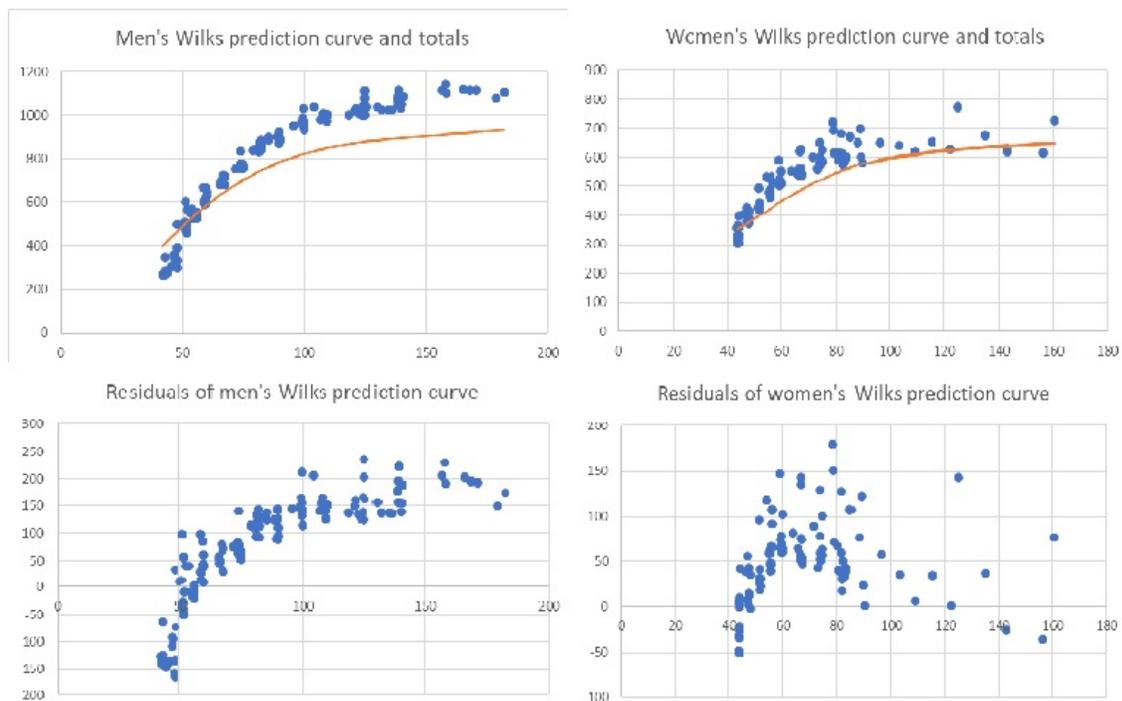

*Figure 1A-D.* The currently used Wilks prediction curves shown along with RAW powerlifting total results, and their residuals (predicted value subtracted from powerlifting total result). The figures show that the current Wilks scoring system is clearly biased, favoring heavyweight men and around 60 kg bodyweight women.

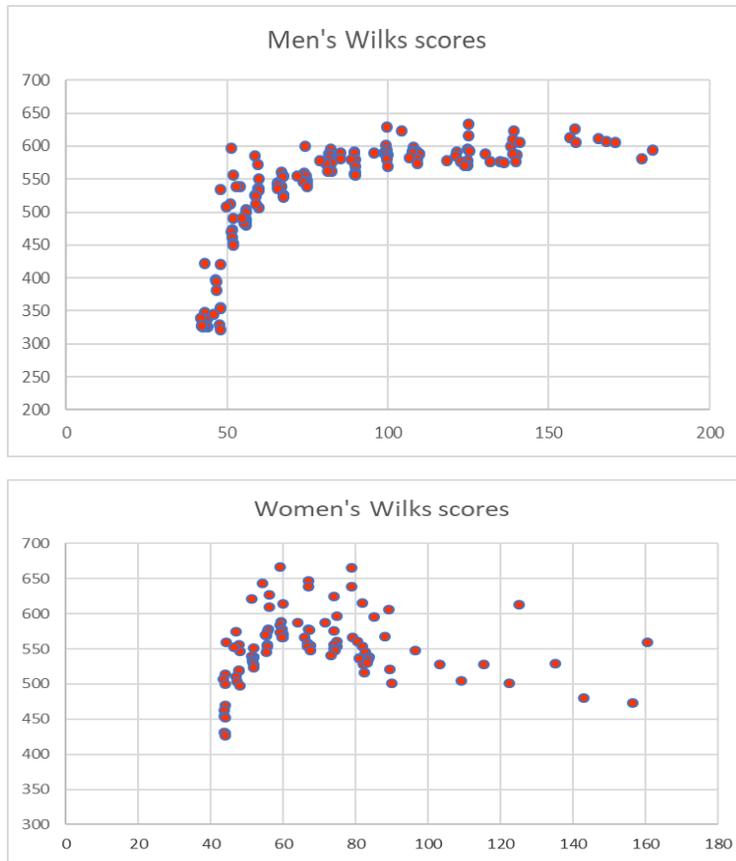

***Figure 2A-B.*** *Wilks score distributions for men and women. An ideal score distribution would be on a horizontal line, and would divide the top 10 ranked scores evenly across the weight classes.*

A 4th order polynomial was chosen instead of a 5th order polynomial like that in the current Wilks system, because the 5th order polynomial produced many plateau areas with the men's data, inconsistent with the physiological nature of the distribution (figures 3A-H). R^2>0.85 in the separate fits for men and women. The residuals of the chosen prediction functions are evenly distributed over the weight classes (figures 4A and 4B). Results of fitting the 4th order polynomials are summarized in Table 1.

**Table 1.** *The 4th order polynomial curve fits and their coefficients of determination R^2.*

f = a + bx + cx^2 + dx^3 + ex^4

|  | a | b | c | d | e | R^2 |
|---|---|---|---|---|---|---|
| Men | 561.53 | -15.807 | 0.47799 | -0.00373 | 9.31 | 0.9628 |
| Women | -898.34 | 48.077 | -0.5618 | 0.00292 | -5.64 | 0.8536 |

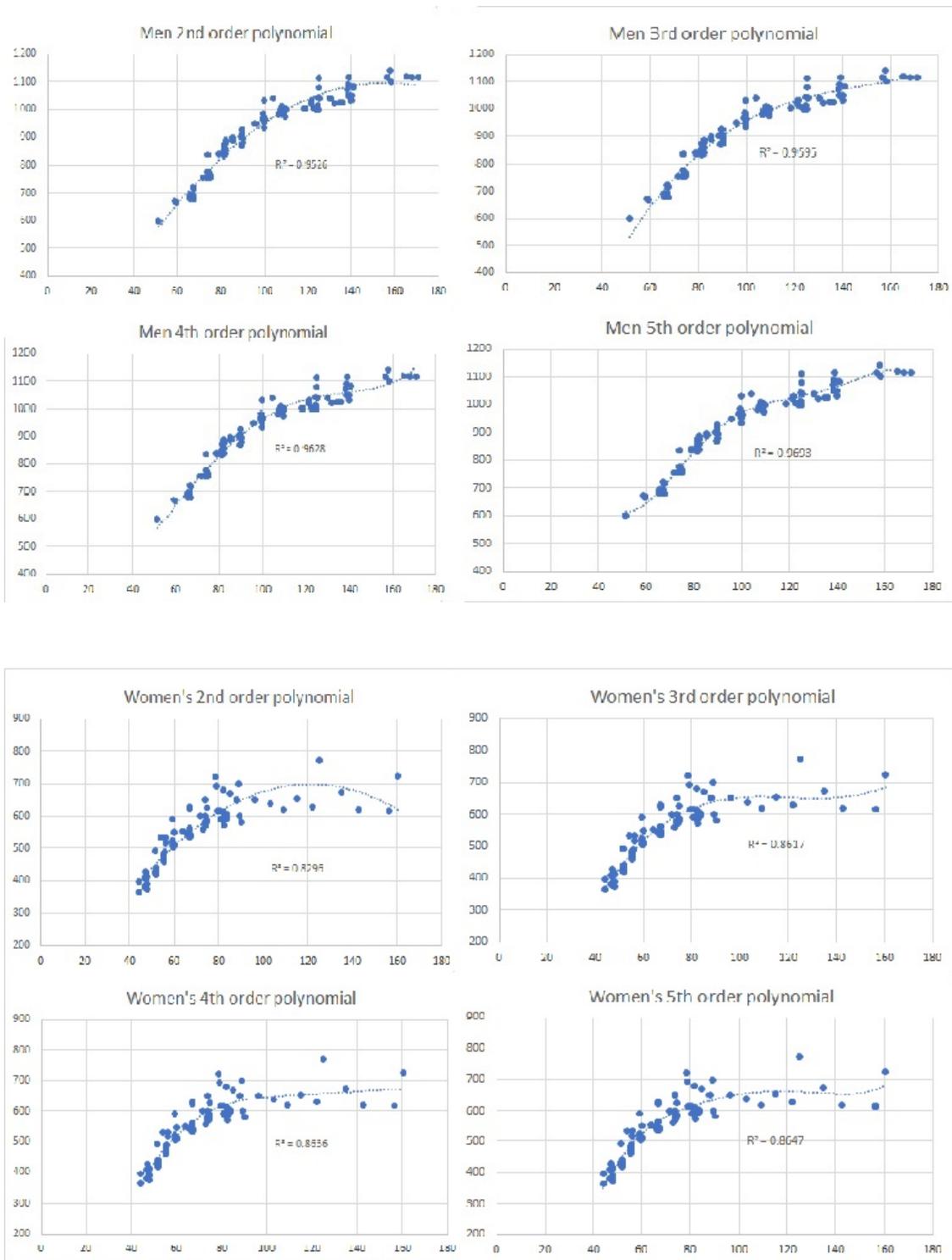

***Figure 3A-H.*** 2$^{nd}$, 3$^{rd}$, 4$^{th}$, and 5$^{th}$ order polynomial fits to men's and women's RAW powerlifting total results. The 2$^{nd}$ order polynomials would produce biased high residuals for heavyweight classes, and 3$^{rd}$ order polynomial for men's lightweight classes. The 5$^{th}$ order polynomial fit has more than one plateau area along men's physiological bodyweight, and thus does not correspond to the physiological dependence of powerlifting results on bodyweight. The 4$^{th}$ order polynomials were chosen for further testing.

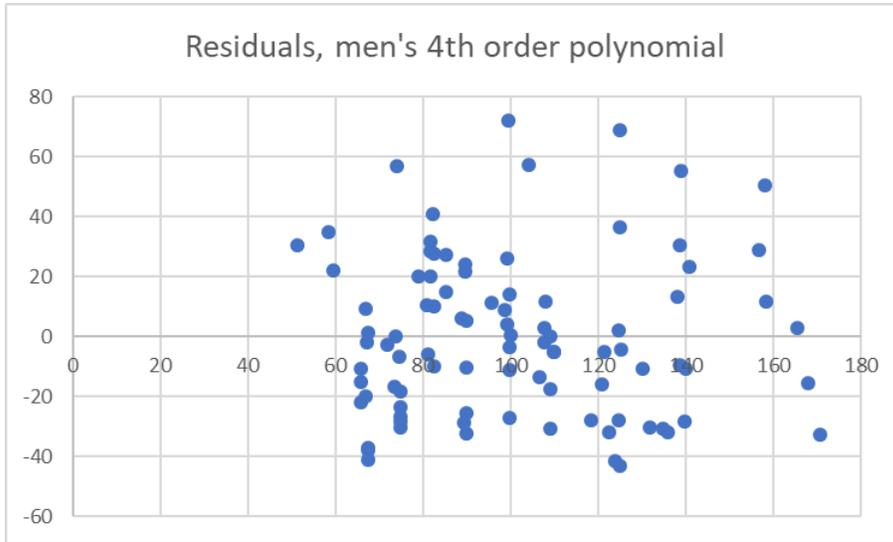

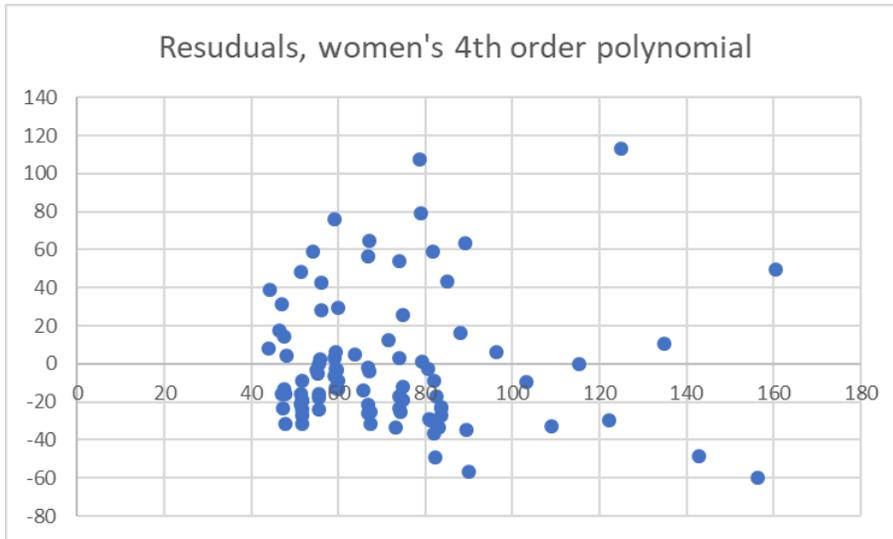

***Picture 4A-B.*** *Residuals of the 4th order polynomial fits show no systematic bias.*

With this dataset, the 4th order polynomials gave fair appearing score distributions for both men and women (figures 5A and 5B). When the omitted weight classes were added to the plots, the results for the very lightweight classes were underestimates (figures 6A and 6B).

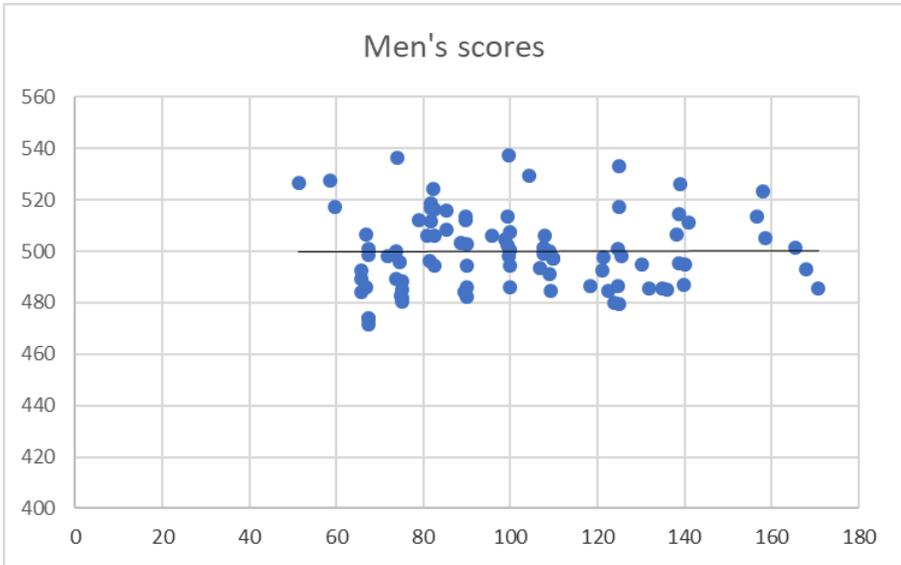

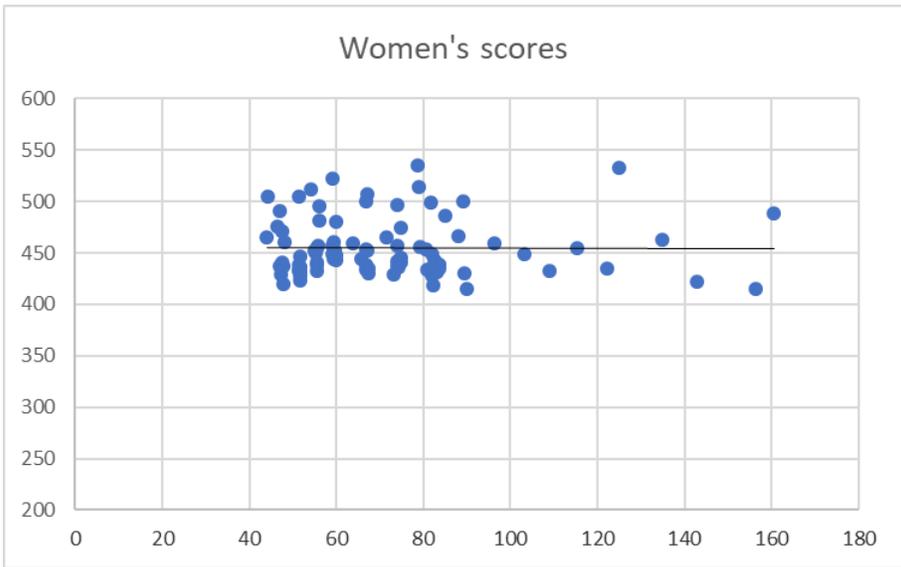

***Picture 5A-B.*** *4th order adjustment functions produce score distributions that appear fair across weight classes for both men and women. The top 10 scores are evenly distributed over the weight classes.*

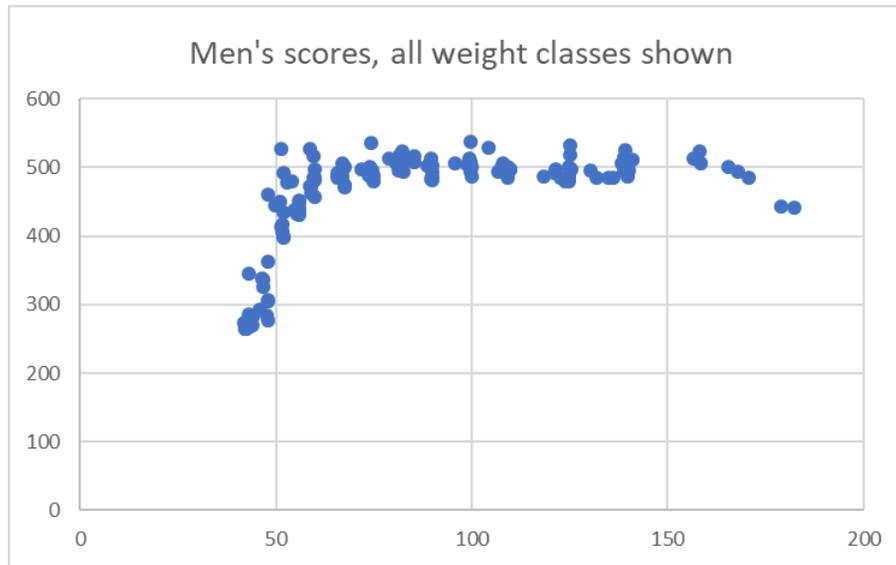

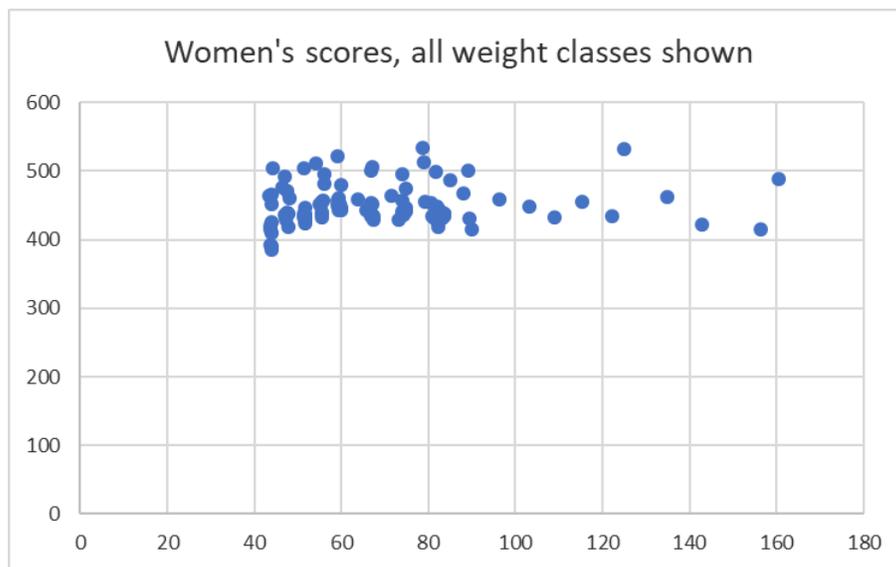

***Picture 6A-B.*** *Full score distributions for men and women. The weight classes that were omitted from fitting the prediction curves are included in these score plots to inspect the full distributions. It is acceptable to underestimate the scores at the extremes of weight, whereas if the scores would be overestimated, the top 10 scores would likely change dramatically. It seems to be acceptable to extrapolate this scoring system down to men's 50 kg bodyweight, and to all women's bodyweights.*

**Conclusions**

Polynomial functions (Wilks formulas) are currently used for comparing pro RAW powerlifting results across bodyweights. This study did not evaluate the method itself, but demonstrated new estimates for the polynomial scoring system, to update the one already used. This study showed that 4$^{th}$ order polynomials were the most efficacious for the current update.

The currently used unrevised Wilks formulas do not fit the current top level RAW powerlifting data. This current scoring system gives clearly biased score distributions.

The 4$^{th}$ order polynomials were fit for adjusting pro level RAW powerlifting total results by body weight, for men with bodyweight in the range 60 – 175 kg, and for women with bodyweight from 44 kg up. It seems acceptable to apply the fits to men with bodyweights from 50 kg up, and to women of all bodyweights. The World's top 10 scores are evenly distributed over the different weight classes for both men and women.

The sample size could not be larger than what was used here, because there are no more than 10 pro level powerlifting athletes in some of the less competed weight classes. Pooling data from individual pro competitions to get a larger sampling would result in using results of the same athlete many times, which should be handled as grouped data, and thus would not yield any larger sample for the actual regression analyses. Scoring systems fitted to larger amateur data samples, like the IPF score, may be statistically unbiased, but do favour medium weight athletes due to the larger number of competitors in these weight classes. Such might be a reasonable scoring system for amateurs, to account for more competition in the middle weight classes, but will result in only middle weight class winners in a pro competition setting.

It is very hard to make a fair comparison between women and men, because the variance of women's results is clearly higher than that for men. This results in more top ranked scores for women than for men. Many competitions compare male and female athletes together using the scores. In this study, men's results were normalized to 500 points and women's to 455 points, along the fitted scoring curves, to bring the top scores to the same level. Normalizing also women's scores to 500 would make the winner of a pro competition always a woman. In this sense, normalizing women's scores lower than men's makes the competition for top scores fairer between the sexes, but leaves the women's medium scores lower than men's because the former scores have larger variance than the latter.

We conclude that the proposed 4$^{th}$ order polynomial scoring system could be considered a candidate for updating the original Wilks formulas, to compare pro RAW powerlifting total results across bodyweights and sexes. The formulas should, in continuation, be assessed/updated every 5 years, since this current assessment clearly demonstrated that deviation from a fit can grow unacceptably as time passes.

**Aknowledgements**

We would like to thank the openpowerlifting.org administration for collecting the huge result data for open access, and keeping the ranking list up to date.

Thank you for professor Seppo Karrila for advice in planning and writing this study.

**Appendix**

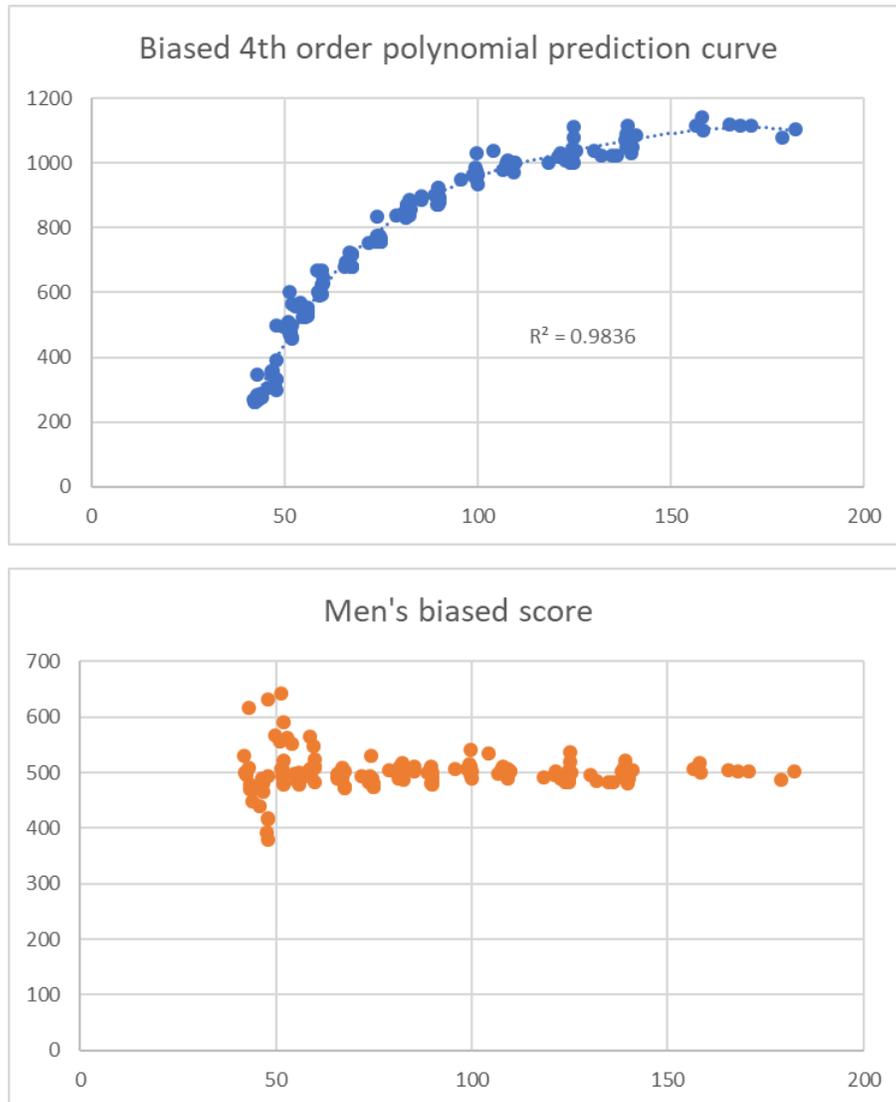

***Picture 7A-B.*** *4th order polynomials fitted to 10 best RAW powerlifting results of all men's weight classes, and score distribution using the same polynomial. The polynomial overestimates the variance of the light weight class results, which makes the score distribution biased, even though the prediction curve seems to have a good fit. Note that the heaviest athlete (around 180 kg bodyweight) got the same score as the 3 lighter heavyweight class athletes (around 170 kg bodyweight), even though the lighter athletes had higher powerlifting totals. This is an example of how a mathematically fit model can produce a biased score system.*